\documentclass[12pt]{article}
\usepackage{lineno}
\usepackage{graphicx}
\usepackage{amsmath}
\usepackage{hhline}
\usepackage{amssymb}
\usepackage{times}
\usepackage{cite}
\usepackage{array}
\usepackage{multirow}
\usepackage{bigstrut}

\usepackage{color}
\definecolor{rltred}{rgb}{0.75,0,0}
\definecolor{rltgreen}{rgb}{0,0.5,0}
\definecolor{rltblue}{rgb}{0,0,0.75}

\newif\ifpdf
\ifx\pdfoutput\undefined
    \pdffalse          
\else
    \pdfoutput=1       
    \pdftrue
\fi

\ifpdf
\linenumbers           
\usepackage{thumbpdf}
\usepackage[pdftex,
        colorlinks=true,
        urlcolor=rltblue,       
        filecolor=rltgreen,     
        linkcolor=rltred,       
        pdftitle={Example File for pdflatex},
        pdfauthor={H1},
        pdfsubject={Template file},
        pdfkeywords={High-Energy Physics, Particle Physics},
        pdfpagemode=None,
        bookmarksopen=true]{hyperref}
 
\fi

\newlength{\dinwidth}
\newlength{\dinmargin}
\begin{document}

\newcommand{\pom}{{I\!\!P}}
\newcommand{\reg}{{I\!\!R}}
\newcommand{\slowpi}{\pi_{\mathit{slow}}}
\newcommand{\fiidiii}{F_2^{D(3)}}
\newcommand{\fiidiiiarg}{\fiidiii\,(\beta,\,Q^2,\,x)}
\newcommand{\n}{1.19\pm 0.06 (stat.) \pm0.07 (syst.)}
\newcommand{\nz}{1.30\pm 0.08 (stat.)^{+0.08}_{-0.14} (syst.)}
\newcommand{\fiidiiiful}{F_2^{D(4)}\,(\beta,\,Q^2,\,x,\,t)}
\newcommand{\fiipom}{\tilde F_2^D}
\newcommand{\ALPHA}{1.10\pm0.03 (stat.) \pm0.04 (syst.)}
\newcommand{\ALPHAZ}{1.15\pm0.04 (stat.)^{+0.04}_{-0.07} (syst.)}
\newcommand{\fiipomarg}{\fiipom\,(\beta,\,Q^2)}
\newcommand{\pomflux}{f_{\pom / p}}
\newcommand{\nxpom}{1.19\pm 0.06 (stat.) \pm0.07 (syst.)}
\newcommand {\gapprox}
   {\raisebox{-0.7ex}{$\stackrel {\textstyle>}{\sim}$}}
\newcommand {\lapprox}
   {\raisebox{-0.7ex}{$\stackrel {\textstyle<}{\sim}$}}
\def\gsim{\,\lower.25ex\hbox{$\scriptstyle\sim$}\kern-1.30ex%
\raise 0.55ex\hbox{$\scriptstyle >$}\,}
\def\lsim{\,\lower.25ex\hbox{$\scriptstyle\sim$}\kern-1.30ex%
\raise 0.55ex\hbox{$\scriptstyle <$}\,}
\newcommand{\pomfluxarg}{f_{\pom / p}\,(x_\pom)}
\newcommand{\dsf}{\mbox{$F_2^{D(3)}$}}
\newcommand{\dsfva}{\mbox{$F_2^{D(3)}(\beta,Q^2,x_{I\!\!P})$}}
\newcommand{\dsfvb}{\mbox{$F_2^{D(3)}(\beta,Q^2,x)$}}
\newcommand{\dsfpom}{$F_2^{I\!\!P}$}
\newcommand{\gap}{\stackrel{>}{\sim}}
\newcommand{\lap}{\stackrel{<}{\sim}}
\newcommand{\fem}{$F_2^{em}$}
\newcommand{\tsnmp}{$\tilde{\sigma}_{NC}(e^{\mp})$}
\newcommand{\tsnm}{$\tilde{\sigma}_{NC}(e^-)$}
\newcommand{\tsnp}{$\tilde{\sigma}_{NC}(e^+)$}
\newcommand{\st}{$\star$}
\newcommand{\sst}{$\star \star$}
\newcommand{\ssst}{$\star \star \star$}
\newcommand{\sssst}{$\star \star \star \star$}
\newcommand{\tw}{\theta_W}
\newcommand{\sw}{\sin{\theta_W}}
\newcommand{\cw}{\cos{\theta_W}}
\newcommand{\sww}{\sin^2{\theta_W}}
\newcommand{\cww}{\cos^2{\theta_W}}
\newcommand{\trm}{m_{\perp}}
\newcommand{\trp}{p_{\perp}}
\newcommand{\trmm}{m_{\perp}^2}
\newcommand{\trpp}{p_{\perp}^2}
\newcommand{\alp}{\alpha_s}

\newcommand{\alps}{\alpha_s}
\newcommand{\sqrts}{$\sqrt{s}$}
\newcommand{\LO}{$O(\alpha_s^0)$}
\newcommand{\Oa}{$O(\alpha_s)$}
\newcommand{\Oaa}{$O(\alpha_s^2)$}
\newcommand{\PT}{p_{\perp}}
\newcommand{\JPSI}{J/\psi}
\newcommand{\sh}{\hat{s}}
\newcommand{\uh}{\hat{u}}
\newcommand{\MP}{m_{J/\psi}}
\newcommand{\PO}{I\!\!P}
\newcommand{\xbj}{x}
\newcommand{\xpom}{x_{\PO}}
\newcommand{\ttbs}{\char'134}
\newcommand{\xpomlo}{3\times10^{-4}}  
\newcommand{\xpomup}{0.05}  
\newcommand{\dgr}{^\circ}
\newcommand{\pbarnt}{\,\mbox{{\rm pb$^{-1}$}}}
\newcommand{\gev}{\,\mbox{GeV}}
\newcommand{\WBoson}{\mbox{$W$}}
\newcommand{\fbarn}{\,\mbox{{\rm fb}}}
\newcommand{\fbarnt}{\,\mbox{{\rm fb$^{-1}$}}}
%
%
\newcommand{\qsq}{\ensuremath{Q^2} }
\newcommand{\gevsq}{\ensuremath{\mathrm{GeV}^2} }
\newcommand{\et}{\ensuremath{E_t^*} }
\newcommand{\rap}{\ensuremath{\eta^*} }
\newcommand{\gp}{\ensuremath{\gamma^*}p }
\newcommand{\dsiget}{\ensuremath{{\rm d}\sigma_{ep}/{\rm d}E_t^*} }
\newcommand{\dsigrap}{\ensuremath{{\rm d}\sigma_{ep}/{\rm d}\eta^*} }
\def\Journal#1#2#3#4{{#1} {\bf #2} (#3) #4}
\def\NCA{\em Nuovo Cimento}
\def\NIM{\em Nucl. Instrum. Methods}
\def\NIMA{{\em Nucl. Instrum. Methods} {\bf A}}
\def\NPB{{\em Nucl. Phys.}   {\bf B}}
\def\PLB{{\em Phys. Lett.}   {\bf B}}
\def\PRL{\em Phys. Rev. Lett.}
\def\PRD{{\em Phys. Rev.}    {\bf D}}
\def\ZPC{{\em Z. Phys.}      {\bf C}}
\def\EJC{{\em Eur. Phys. J.} {\bf C}}
\def\CPC{\em Comp. Phys. Commun.}

\begin{titlepage}

\begin{flushleft}
DESY 04-209 \hfill ISSN 0418-9833 \\
October 2004
\end{flushleft}

\vspace{2cm}

\begin{center}
\begin{Large}
  
   {\bf Measurement of {\boldmath $F_2^{c\bar{c}}$} and 
{\boldmath $F_2^{b\bar{b}}$}
      at High {\boldmath $Q^2$} \\ using the H1 Vertex Detector at HERA}

\vspace{2cm}

H1 Collaboration

\end{Large}
\end{center}

\vspace{2cm}

\begin{abstract}
  \noindent
  Measurements are presented of inclusive charm and beauty cross sections in
  $e^+p$ collisions at HERA for values of photon virtuality 
  $Q^2 > 150~{\rm GeV}^2$ and of inelasticity 
  $0.1 < y < 0.7$. \linebreak The  charm and beauty fractions
  are determined using a
  method based on the impact parameter, in the 
  transverse plane, of tracks to the 
  primary vertex, as measured by the H1 vertex detector.  The data are
  divided into four regions in $Q^2$ and Bjorken $x$, and values for the
  structure functions $F_2^{c\bar{c}}$ and $F_2^{b\bar{b}}$ are
  obtained.  The results are found to be compatible with the
  predictions of perturbative quantum chromodynamics.  

\end{abstract}

\vspace{1.5cm}

\begin{center}
To be submitted to {\em Eur. Phys. J.} {\bf C}
\end{center}

\end{titlepage}

\begin{flushleft}

A.~Aktas$^{10}$,               
V.~Andreev$^{26}$,             
T.~Anthonis$^{4}$,             
S.~Aplin$^{10}$,               
A.~Asmone$^{33}$,              
A.~Babaev$^{25}$,              
S.~Backovic$^{37}$,            
J.~B\"ahr$^{37}$,              
A.~Baghdasaryan$^{36}$,        
P.~Baranov$^{26}$
% ...

\bigskip{\it
\noindent
% ...
$ ^{4}$ Inter-University Institute for High Energies ULB-VUB, Brussels;
          Universiteit Antwerpen, Antwerpen; Belgium$^{ c}$ \\
% ...
$ ^{10}$ DESY, Hamburg, Germany \\
% ...
$ ^{25}$ Institute for Theoretical and Experimental Physics,
          Moscow, Russia$^{ l}$ \\
 $ ^{26}$ Lebedev Physical Institute, Moscow, Russia$^{ e}$ \\
% ...
$ ^{33}$ Dipartimento di Fisica Universit\`a di Roma Tre
          and INFN Roma~3, Roma, Italy \\
% ...
$ ^{36}$ Yerevan Physics Institute, Yerevan, Armenia \\
 $ ^{37}$ DESY, Zeuthen, Germany \\
% ...
$ ^b$ Supported by the UK Particle Physics and Astronomy Research
      Council, and formerly by the UK Science and Engineering Research
      Council \\
 $ ^c$ Supported by FNRS-FWO-Vlaanderen, IISN-IIKW and IWT
      and  by Interuniversity
Attraction Poles Programme,
      Belgian Science Policy \\
% ...
$ ^e$ Supported by the Deutsche Forschungsgemeinschaft \\
% ...
$ ^l$ Partially Supported by Russian Foundation
      for Basic Research, grant    no. 00-15-96584 \\
}

\end{flushleft}

\newpage

\section{Introduction}
Heavy quark production is an important process to study quantum
chromodynamics (QCD). It is expected that perturbative QCD (pQCD) at 
next-to-leading order (NLO) should give a good description of heavy flavour
production in deep-inelastic scattering (DIS), especially at values of
the negative square of the four momentum of the exchanged boson $Q^2$
greater than the square of the heavy quark masses.  Measurements of the
open charm ($c$) cross section in DIS at HERA have mainly been of exclusive
$D$ or $D^*$ meson production\cite{H1ZEUSDstar,Chekanov:2003rb}.
From these measurements the contribution of charm to the proton
structure function, $F_2^{c\bar{c}}$, is derived by correcting for the
fragmentation fraction $f(c \rightarrow D)$ and the unmeasured phase
space (mainly at low values of transverse momentum of the meson). The
results are found to be in good agreement with pQCD predictions.  The
measurement of the beauty ($b$) cross section is particularly
challenging since $b$ events comprise only a small fraction (typically
$< 5\%$) of the total cross section. The $b$ cross section has been
measured in DIS ($Q^2 > 2~{\rm GeV^2}$) by ZEUS\cite{zeusBdis} and in
photoproduction ($Q^2 \simeq 0~{\rm GeV^2}$) by H1\cite{Adloff:1999nr}
and ZEUS\cite{zeusBgammap}, using the transverse momentum distribution
of muons relative to the $b$ jet in semi-muonic decays.  Measurements
of the $b$ cross section at high centre of mass energy have also been
made in $p \bar{p}$\cite{hadronb} and $\gamma \gamma$
collisions\cite{ggb}.

The analysis presented in this paper is of inclusive $c$ and
$b$ cross sections in $e^+p$ scattering at HERA in the 
range $Q^2 > 150~{\rm GeV}^2$.
Events containing heavy quarks can be distinguished from light quark
events by the long lifetimes of $c$ and $b$ flavoured hadrons,
which lead to displacements of tracks from the primary vertex. 
The distance of a track to the primary vertex is reconstructed using 
precise spatial information from the H1 vertex 
detector.
The results presented in this paper are
made in kinematic regions where 
there is little extrapolation needed to correct to the
full phase space  and so the model dependent uncertainty due to the
 extrapolation is small.
The charm structure function $F_2^{c\bar{c}}$ and the corresponding
structure function for $b$ quarks $F_2^{b\bar{b}}$ are obtained after
small corrections for the longitudinal structure functions
$F_L^{c\bar{c}}$ and $F_L^{b\bar{b}}$.  This is an extension to high
$Q^2$ of previous H1 $F_2^{c\bar{c}}$ measurements and the first
measurement of $F_2^{b\bar{b}}$.

\section{Theory of Heavy Flavour Production in DIS}
\label{sec:theory}
In pQCD, in the region where $Q^2$ is much larger than the squared
mass $M^2$ of the heavy quark, the production of heavy flavour
quarks is expected to be insensitive to threshold effects and the
quarks may be treated as massless partons.
At leading order (LO), in the `massless' scheme, the quark parton
model (QPM) process ($\gamma q \rightarrow q$) is the dominant
contribution. At NLO, the photon gluon fusion ($\gamma g \rightarrow
q\bar{q}$) and QCD Compton ($\gamma q \rightarrow qg$) processes also
contribute.
The approach is often referred to as the zero mass variable flavour
number scheme (ZM-VFNS)\cite{cteq4,Martin:1994kn}.

At values of $Q^2 \sim M^2$,
the `massive' scheme\cite{massive}, in which the heavy flavour partons
are treated as massive quarks is more appropriate. 
  The dominant LO process is photon
gluon fusion (PGF) and the NLO diagrams are of order $\alpha_s^2$.
The scheme is often referred to as the fixed flavour number scheme
(FFNS).  As $Q^2$ becomes large compared to $M^2$, the FFNS approach is
unreliable due to large logarithms in $\ln (Q^2/M^2)$ in the perturbative
series.

In order to provide reliable pQCD predictions for the description of
heavy flavour production, over the whole range in $Q^2$, composite
schemes which provide a smooth transition from the massive description
at $Q^2 \sim M^2$ to the massless behaviour at $Q^2 \gg M^2$ have been
developed\cite{VFNS1,VFNS2}.
The scheme is commonly referred to as the variable flavour number
scheme (VFNS).
The approach has been incorporated in various different forms to order
$\alpha_s$\cite{VFNS1}
and to order $\alpha_s^2$\cite{VFNS2}.

\section{H1 Detector}
Only a short description of the H1 detector is given here; a full
description may be found in\cite{Abt:1997xv}. A right handed
coordinate system is employed at H1 that has its $z$-axis pointing in
the proton beam, or forward, direction and $x$ ($y$) pointing in
the horizontal (vertical) direction.

Charged particles are measured in the central tracking detector (CTD).
This device consists of two cylindrical drift chambers interspersed with
$z$-chambers to improve the $z$-coordinate reconstruction and
multi--wire proportional chambers mainly used for triggering. The CTD
is situated in a uniform $1.15\,{\rm T}$ magnetic field, enabling
momentum measurement of charged particles over the polar angular
range $20^\circ< \theta<160^\circ$~\footnote{\noindent{ The angular coverage of each detector 
component is given for the interaction
  vertex in its nominal position.}}.  

The CTD tracks are linked to hits in the vertex detector (central
silicon tracker CST)\cite{cst}, to provide precise spatial track
reconstruction. The CST consists of two layers of double-sided silicon
strip detectors surrounding the beam pipe, covering an angular range
of $30^\circ < \theta< 150^\circ$ \linebreak for tracks passing through both
layers. This detector provides hit resolutions of $12$~$\mu$m in
 \linebreak $r$--$\phi$ and $25$~$\mu$m in $z$. For CTD tracks with CST hits in
both layers the transverse distance of closest approach (DCA) to the nominal
vertex in $x$--$y$ can be measured with a resolution of 
$33\;\mu\mbox{m} \oplus 90 \;\mu\mbox{m} /p_T [\mbox{GeV}]$, where the
first term represents the intrinsic resolution (including alignment
uncertainty) and the second term is the contribution from multiple
scattering in the beam pipe and the CST; $p_T$ is the transverse
momentum of the track.

The track detectors are surrounded in the forward and central
directions ($4^\circ<\theta<155^\circ$) by a fine grained liquid argon
calorimeter (LAr) and in the backward region
($153^\circ<\theta<178^\circ$) \linebreak
by a lead--scintillating fibre
calorimeter\cite{Nicholls:1996di} with electromagnetic and hadronic
sections. These calorimeters provide energy and angular reconstruction
for final state particles from the hadronic system. The LAr is also
used in this analysis to measure and identify the scattered positron.
An electromagnetic calorimeter situated downstream in the positron
beam direction measures photons from the bremsstrahlung
process $ep\rightarrow ep\gamma$ for the purpose of luminosity
determination.

\section{Experimental Method}
The analysis is based on a high $Q^2$ sample of $e^+p$ neutral current
(NC) scattering events corresponding to an integrated luminosity of
$57.4$ ${\rm pb}^{-1}$, taken in the years 1999-2000, at an $ep$
centre of mass energy $\sqrt{s} = 319~{\rm GeV}$.  The events are
selected as described in\cite{H19900NCCC}; the positron is identified
and measured in the LAr calorimeter, which restricts the sample
to $Q^2>110~{\rm GeV}^2$.  The event kinematics, $Q^2$ and the
inelasticity variable $y$, are reconstructed using
the scattered positron.  The Bjorken scaling variable $x$
is obtained from $x = Q^2/sy$. After the inclusive selection 
the total number of events is around $121,000$.

\subsection{Monte Carlo Simulation}
The data are corrected for the effects of detector resolution,
acceptance and efficiency by the use of Monte Carlo simulations.
The Monte Carlo program RAPGAP\cite{Jung:1993gf} is used to generate
high $Q^2$ NC DIS events for the processes 
$ep \rightarrow eb\bar{b}X$, $ep \rightarrow ec\bar{c}X$ and
light quark production.
RAPGAP combines  $\cal{O}$($\alpha_s$) matrix
elements with higher order QCD effects modelled by the emission of
parton showers. The heavy flavour event samples are generated
according to the massive PGF matrix element with the mass of the $c$
and $b$ quarks set to $m_c=1.5 \ {\rm GeV}$ and $m_b=5.0 \ {\rm GeV}$,
respectively.  The partonic system is fragmented according to the LUND
string model implemented within the JETSET
program\cite{Sjostrand:1993yb}.  The HERACLES
program\cite{Kwiatkowski:1990es}  calculates single photon
radiative emissions off the lepton line and virtual electroweak
corrections.
In the event generation, the DIS cross section is calculated with a LO
parton distribution function (PDF)\cite{Martin:1994kn}.
In order to improve the description of the data by the simulation, the
simulated inclusive cross section is reweighted in $x$ and $Q^2$ using a NLO QCD 
fit (H1 PDF 2000) to the H1 data\cite{H19900NCCC}.

The samples of events generated for the $uds$, $c$ and $b$ 
processes are passed through a detailed simulation of the detector 
response based on the GEANT3
program\cite{Brun:1978fy}, and through the same reconstruction
software as is used for the data.

\subsection{Track, Vertex and Jet Reconstruction}
\label{sec:trackjetvertex}
The analysis is based on CTD tracks which are linked to $r$--$\phi$
hits in both planes of the CST in order to improve the precision of
the track parameters.  In this paper, the CST-improved CTD tracks are
referred to as `CST tracks'.  Only those events which have at least one
reconstructed CST track with polar angle $30^\circ<\theta_{\rm
  track}<150^\circ$ and a minimum transverse momentum of 
$0.5 \ {\rm GeV}$ are used.   At low values of $y$, the hadronic
final state (HFS) tends to go forward and outside the acceptance of
the CST.  Therefore, the analysis is restricted to $0.07<y<0.7$.  The
upper $y$ cut ensures a good trigger acceptance for the scattered
positron.  In this kinematic range, studies from Monte Carlo
simulations show that $93\%$ of $c$ events and $96\%$ of $b$ events
are expected to have at least one charged particle, with $p_T>0.5 \ 
{\rm GeV}$ in the angular range $30^\circ<\theta <150^\circ$, produced from
the decay of a $c$ or $b$ hadron.  The extrapolation to the full phase
space, needed to calculate $F_2^{c\bar{c}}$ and $F_2^{b\bar{b}}$, is
therefore small.

\begin{figure}[htb]
  \begin{center}
    \includegraphics[width=1.0\textwidth]{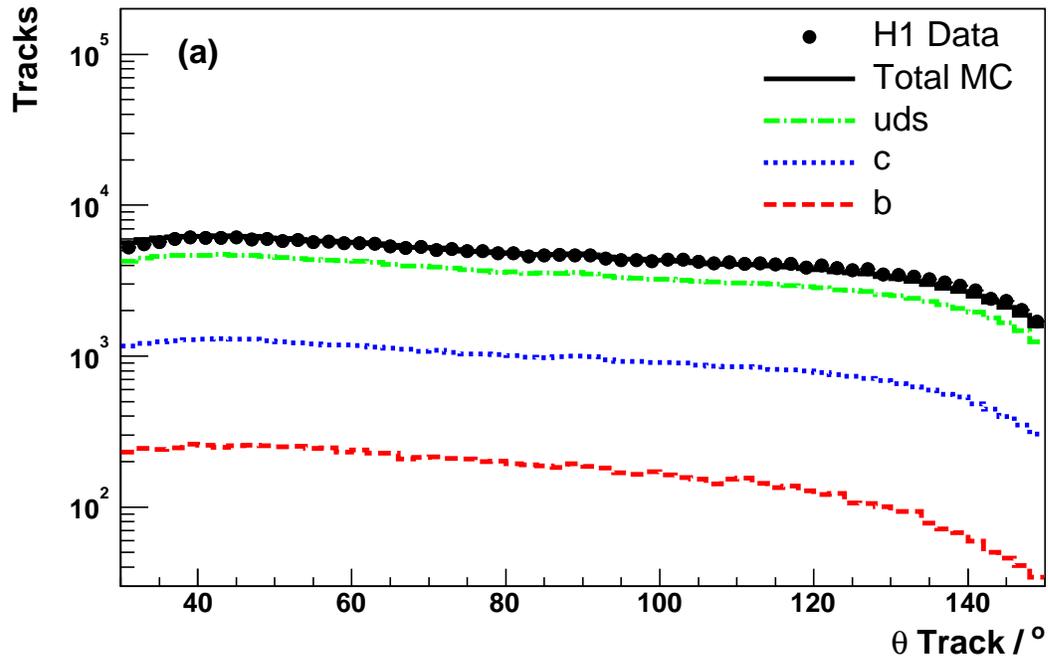}
  \includegraphics[width=1.0\textwidth]{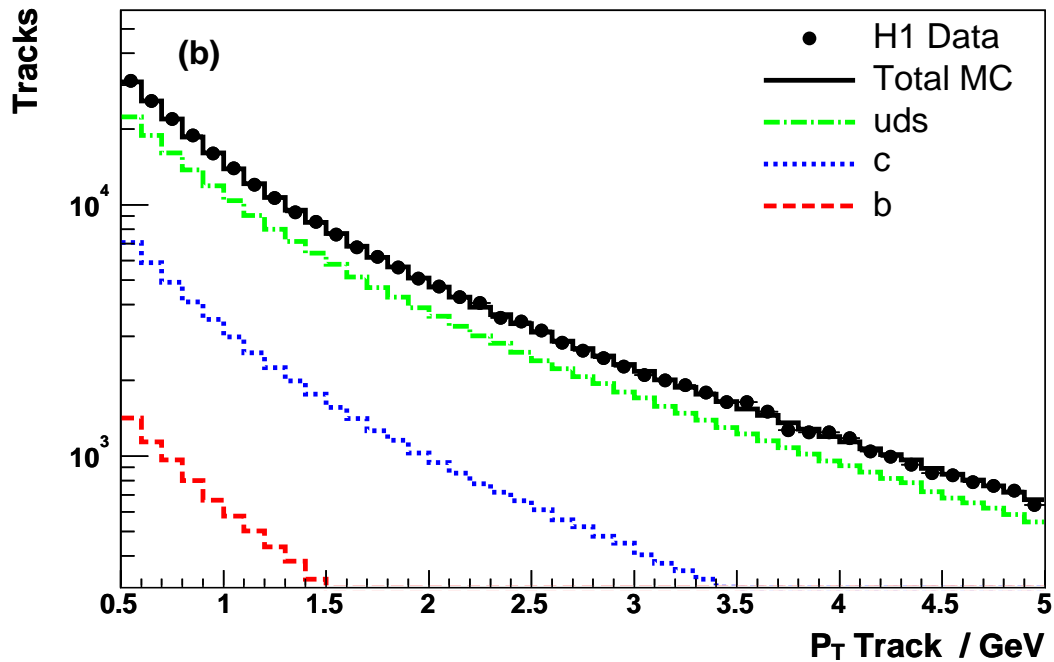}

    \caption{The polar angle distribution (a) and transverse momentum
    distribution (b) of all HFS CST tracks.  Included in the figure is the
    expectation from the RAPGAP Monte Carlo simulation, showing the
    contributions from the various quark flavours after applying the
    scale factors obtained from the fit to the subtracted significance
    distributions of the data (see section~\ref{sec:fit}).}
    \label{fig:thetatracks} \end{center}
\end{figure}

The reconstructed $z$ position of the interaction vertex must be
within $\pm 20~{\rm cm}$ of the centre of the detector to match the
acceptance of the CST.  The effect of the smearing of the $z$-vertex
distribution around the nominal position 
reduces the number of selected events by $\sim 5$\%. 
The CST track reconstruction efficiency is $71$\% for a single charged particle with
$p_T>0.5$~${\rm GeV}$ 
that passes through the CST acceptance region.
This efficiency
includes the CST hit efficiency, CST-CTD linking efficiency and 
losses due to inactive CST regions.  
The polar angle and transverse momentum distributions of
HFS CST tracks are compared to the Monte Carlo simulation in
figure~\ref{fig:thetatracks}. The simulation gives a reasonable
description of these distributions.

The primary event vertex in $r$--$\phi$ is reconstructed from all
tracks (with or without CST hits) and the position and spread of the
beam interaction region (referred to as the `beam-spot'). The
beam-spot extension is measured to be $\sim 145~{\rm \mu m}$ in $x$
and $\sim 25~{\rm \mu m}$ in $y$ for the data period considered here.
The position of the beam-spot
is measured as the average over many events and the
resulting error on the position is small in comparison to the size of
the beam-spot, with a typical uncertainty of $\sim 5~{\rm \mu m}$.
The uncertainty on the primary event vertex for the kinematic
range of this paper is on average $50~{\rm \mu m}$ in $x$ and  
$24~{\rm \mu m}$ in $y$.

In this analysis the impact parameter, i.e. the transverse distance of
closest approach of the track to the primary vertex point, is used to
separate the different quark flavours (see section
\ref{quarkflavourseparation}).  The uncertainty of the measurement of
the impact parameter receives contributions from the position of the
primary vertex discussed above, the intrinsic resolution of the track
and distortions due to multiple scattering in the beam-pipe and
surrounding material.  In order to provide a successful description of
the data the Monte Carlo parameters for the beam-spot size, tracking
resolution and detector material are adjusted to those observed in the
data.

To identify long lived hadrons a `jet axis' is defined 
for each event 
in order to calculate a signed impact parameter ($\delta$) for each track. 
Jets with a minimum $p_T$ of $5 \ {\rm GeV}$,
in the angular range $10^\circ < \theta < 170^{\rm o}$, are reconstructed
using the invariant $k_T$ algorithm\cite{KTJET} in the laboratory
frame using all reconstructed HFS particles.  HFS particles are
reconstructed using a combination of tracks and calorimeter energy
deposits\cite{Adloff:1997mi}. The jet axis is defined as the direction
of the jet with the highest transverse momentum or, if there is no jet
reconstructed in the event, as the `direction of the struck quark in the
quark parton model'\cite{Ahmed:1992sk} as 
reconstructed from the HFS particles.  
In the $Q^2$ range of this
paper, the vector sum of all HFS particles in the laboratory frame always 
has a transverse momentum greater than $5~{\rm GeV}$ and 
$97\%$ of the events have the jet axis defined by a
reconstructed jet.

\begin{figure}[htb]
  \begin{center}
    \includegraphics[width=0.95\textwidth]{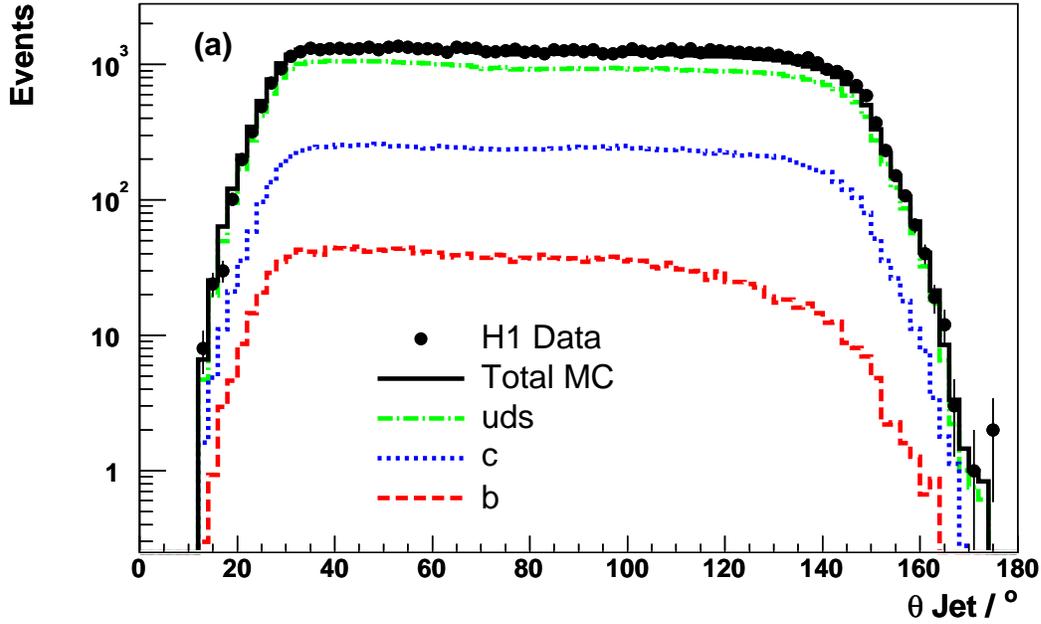}
    \includegraphics[width=0.95\textwidth]{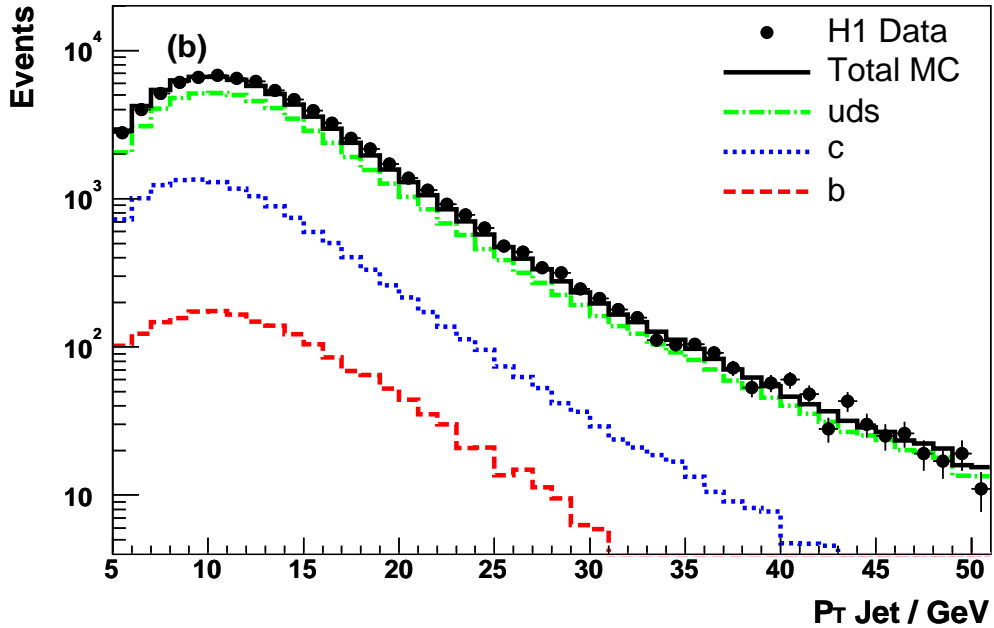}
    \caption{The polar angle distribution (a) and transverse momentum 
      distribution (b) of the highest $p_T$ jet
      which contains at least one
      reconstructed CST track within a cone of radius 1.  If there are
      no reconstructed jets the complete hadronic final state is
      used to define the jet axis. Included in the figure is the expectation from
      the 
      RAPGAP Monte Carlo simulation
      showing the
      contributions from the various quark flavours after
      applying the scale factors obtained from the fit to the
      subtracted significance distributions of the data.}  
    \label{fig:jettheta}
  \end{center}
\end{figure}

\begin{figure}[htb]
  \begin{center} \includegraphics[width=1.0\textwidth]{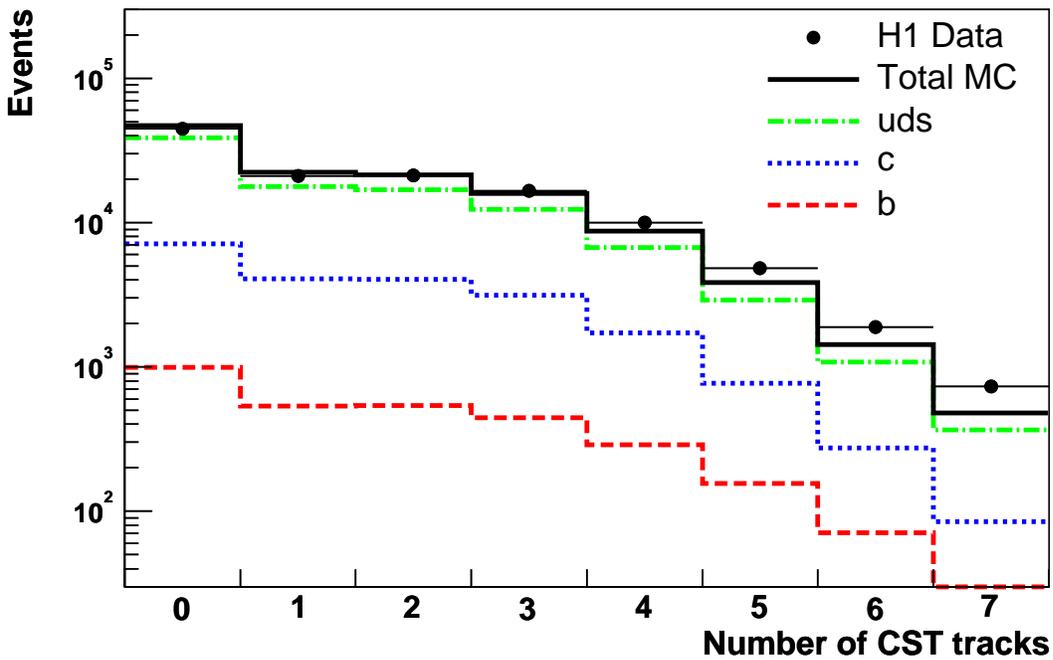}
  \caption{The number of reconstructed central silicon tracker (CST)
  tracks per event associated to the jet axis. Each CST track is required to 
  have at least two CST hits
  and $p_T>0.5~{\rm GeV}$. Included in the figure is the expectation from the RAPGAP
  Monte Carlo simulation, showing the contributions from the various quark
  flavours after applying the scale factors obtained from the fit to the
  subtracted significance distributions of the data.}
\label{fig:csttracks} 
  \end{center}

\end{figure}

CST tracks are associated to the jet axis if they lie within a cone of
size $1$ in pseudo-rapidity--$\phi$ space centred about the jet axis.
Approximately $90\%$ of those events with at least one HFS CST track
have at least one CST track matched to the jet axis.
Figure~\ref{fig:jettheta} shows the polar angle and $p_T$
distributions of the jets which contain one or more CST tracks.
Figure~\ref{fig:csttracks} shows the number of reconstructed CST
tracks associated to the jet axis.  The simulation gives a reasonable
description of these distributions apart from at high multiplicities
where the Monte Carlo is seen to lie a little below the data. The
deviations are due to a non-perfect modelling of multiplicities in
light quark jets and have a negligible effect on the measurements.
The uncertainties on the heavy quark multiplicities and modelling are
discussed in section~\ref{sec:systematics}.

\subsection{Quark Flavour Separation}
\label{quarkflavourseparation}

The different quark flavours that contribute to the DIS cross section
are distinguished on the basis of the different lifetimes of the
produced hadrons. Due to the relatively low cross sections and limited
CST track reconstruction efficiency the decay length of the heavy
hadrons is not reconstructed directly, but the impact parameter of
tracks is used instead.  The results, however, are checked by using an
independent method based on the reconstruction of a secondary vertex
(see section~\ref{sec:secvertex}). The chosen heavy flavour tagging
method also allows events with only one CST track to be used, for
which it is not possible to reconstruct a secondary vertex. For tracks 
associated to the jet axis, $\delta$ is
defined as positive if the angle between the jet axis and the line
joining the primary vertex to the point of DCA is less than $90^\circ$, and is
defined as negative otherwise. 
Tracks from the decays of long lived particles will mainly have a
positive $\delta$.  Tracks produced at the primary vertex result in a
symmetric distribution around $\delta=0$, i.e. 
negative $\delta$ tracks mainly result from detector resolution.

\begin{figure}[htb]
  \begin{center} \includegraphics[width=0.95\textwidth]{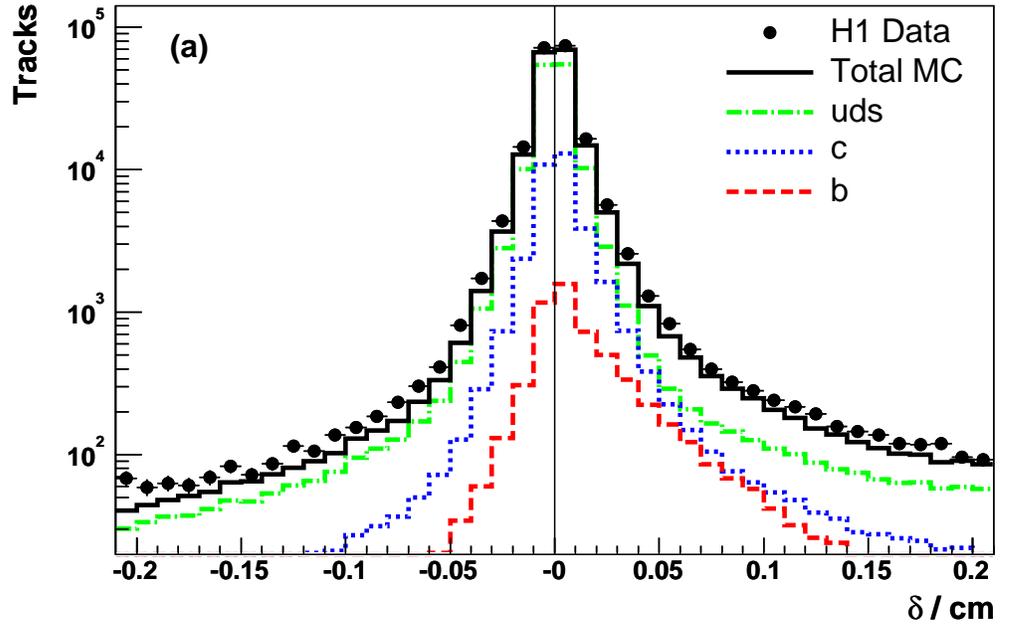}
    \includegraphics[width=0.95\textwidth]{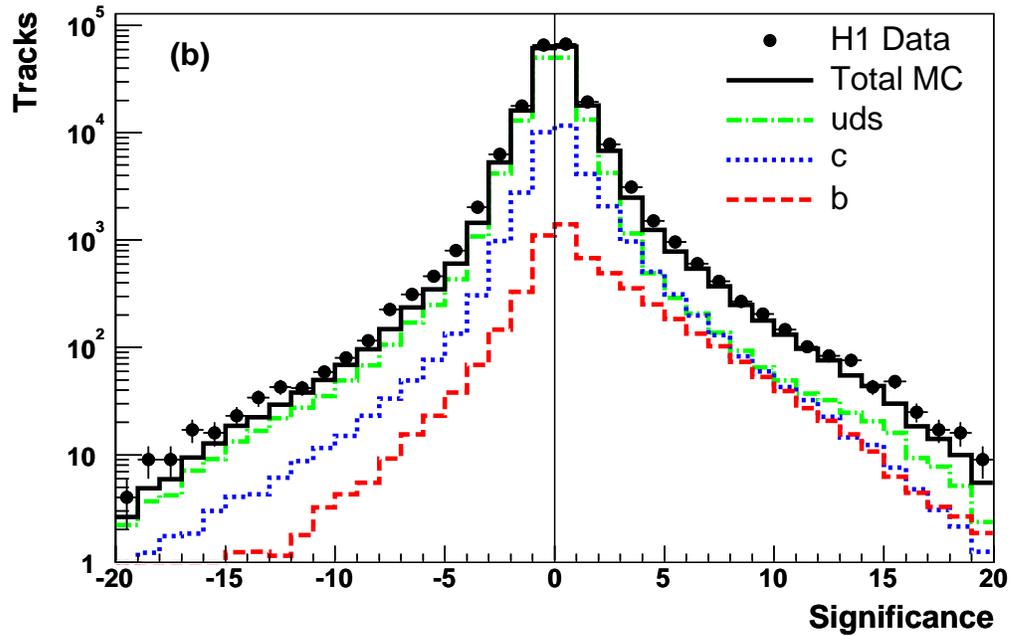} \caption{The signed
      impact parameter $\delta$ of a track to the primary vertex in the
      $x$--$y$ plane (a) and the significance $\delta / \sigma (\delta)$
      (b), where $\sigma(\delta)$ is the error on $\delta$, for all CST tracks
      associated to the jet axis.  The cut $|\delta| < 0.1$~cm has
      been applied in figure (b).  Included in the figure is the
      expectation from the RAPGAP Monte Carlo simulation, showing the contributions
      from the various quark flavours after applying the scale factors
      obtained from the fit to the subtracted significance distributions
      of the data.}  
    \label{fig:dca} 
  \end{center}
\end{figure}

Figure~\ref{fig:dca}(a) shows the $\delta$ distribution of CST tracks
associated to the jet axis.  The data are seen to be asymmetric with
positive values in excess of negative values indicating the presence
of long lived particles. The simulation gives a reasonable
description of the data. The component of the simulation that arises
from light quarks is almost symmetric at low $\delta$.  The
asymmetry at $\delta~\gapprox ~0.1~{\rm cm}$ is mainly due to long
lived strange particles such as $K^0_S$. 
The $c$ component exhibits
a moderate asymmetry and the $b$ component shows a marked
asymmetry. The differences are due to the different lifetimes of the
produced hadrons.  In order to reduce the effects of the strange
component, a cut of $|\delta|< 0.1~{\rm cm}$ is imposed on all
tracks used in the analysis.

In order to optimise the separation of the quark flavours use is made
of the significance, defined as the ratio of
$\delta$ to its error. This distribution is shown for all tracks
in figure~\ref{fig:dca}(b), where a good description of the data by
the simulation is observed apart from the tails. In the tails the data are
observed to lie above the simulation, which is likely to be due to a
non-perfect description of the resolution by the simulation.
The differences in resolution between data and
simulation are treated as a systematic error (see
section~\ref{sec:systematics}).  

\begin{figure}[htb]
  \begin{center}
    \includegraphics[width=0.95\textwidth]{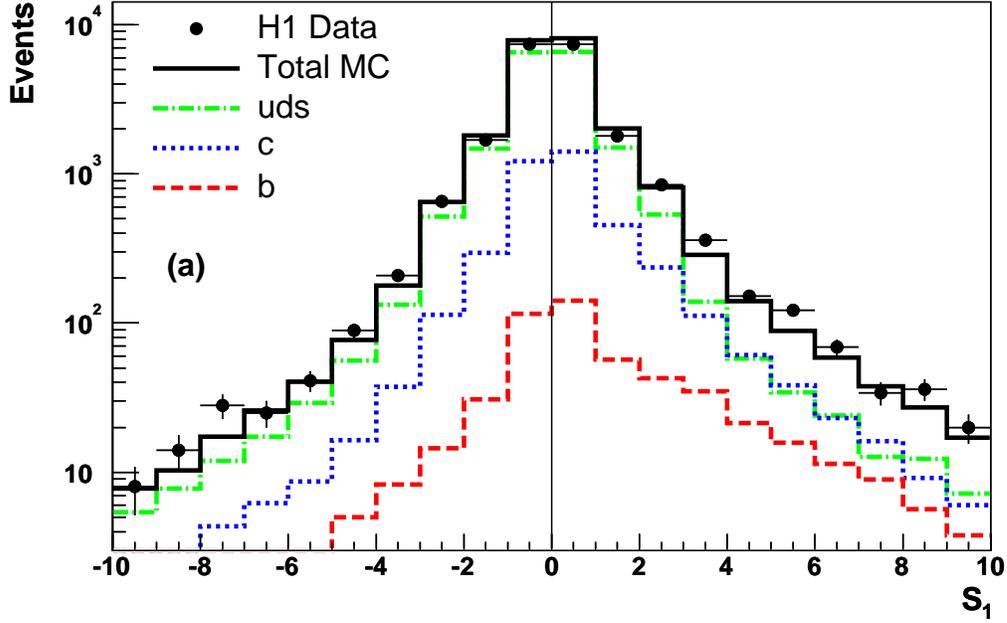}
    \includegraphics[width=0.95\textwidth]{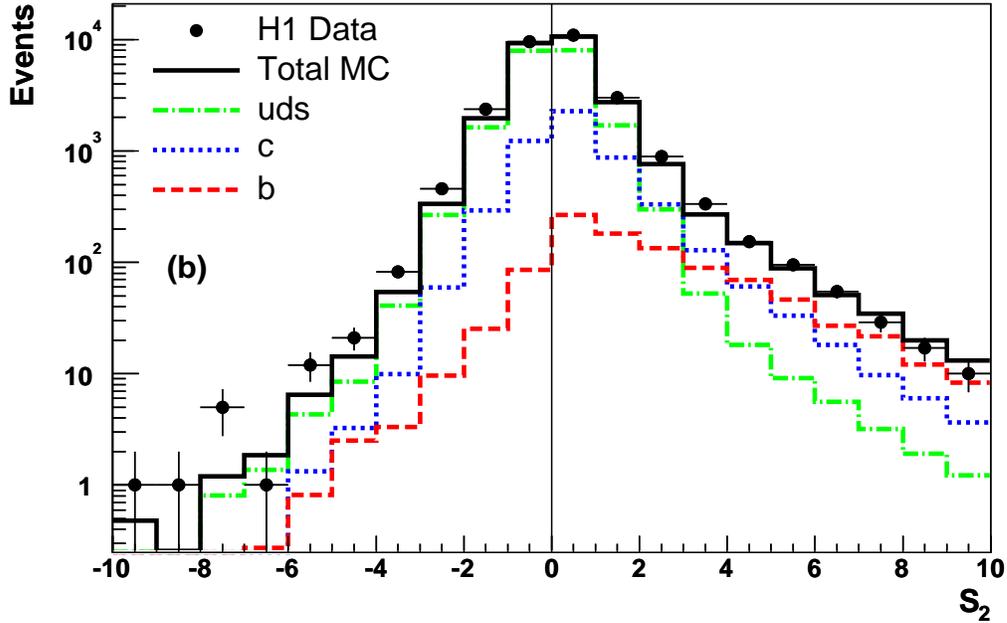}
    \caption{The significance $S_1=$   $\delta /\sigma(\delta)$ 
      distribution per event (a)  for events that contain 
       one reconstructed CST track associated to the jet axis and the 
      significance $S_2= \delta /\sigma (\delta)$  distribution per event 
      (b) of the track with the second highest absolute significance 
      for events with $\ge 2$ reconstructed CST tracks associated to the jet.
      Included in the figure is
      the expectation from the RAPGAP Monte Carlo simulation, showing the
      contributions from the various quark flavours after applying the
      scale factors obtained from the fit to the subtracted significance
      distributions of the data.}
    \label{fig:s1s2} 
  \end{center}
\end{figure}

A further optimisation is made by
using different significance distributions for events with different
multiplicities.  The first significance distribution $S_1$ is defined
for events where only one reconstructed CST track is linked to the jet,
and is simply the significance of the track.  The second significance
distribution $S_2$ is defined for events with two or more tracks
associated with the jet and is the significance of the track with the
second highest absolute significance.
Only events in which the tracks with the first and second
highest absolute significance have the same sign are selected
for the $S_2$ distribution.
The second highest significance track is chosen because for heavy
quarks $\ge 2$ tracks are usually produced with high significance,
whereas for light quarks the chances are small of two tracks being
produced at large significance due to resolution effects.
The $S_1$ and $S_2$ distributions are shown in figure~\ref{fig:s1s2}.
The distribution of $S_2$
gives a better separation power of light to heavy quarks.
Events with one CST track are
retained to improve the statistical precision of the measurements.

In order to substantially reduce the uncertainty due to the resolution
of $\delta$ and the light quark normalisation the negative bins in the
$S_1$ and $S_2$ distributions are subtracted from the positive. The
subtracted distributions are shown in figure~\ref{fig:s1negsub}. It
can be seen that the resulting distributions are dominated by $c$
quark events, with an increasing $b$ fraction with increasing
significance. The light quarks contribute a small fraction for all
values of significance.

\subsection{Fit Procedure}
\label{sec:fit}
\begin{figure}[htb]
  \begin{center} \includegraphics[width=1.0\textwidth]{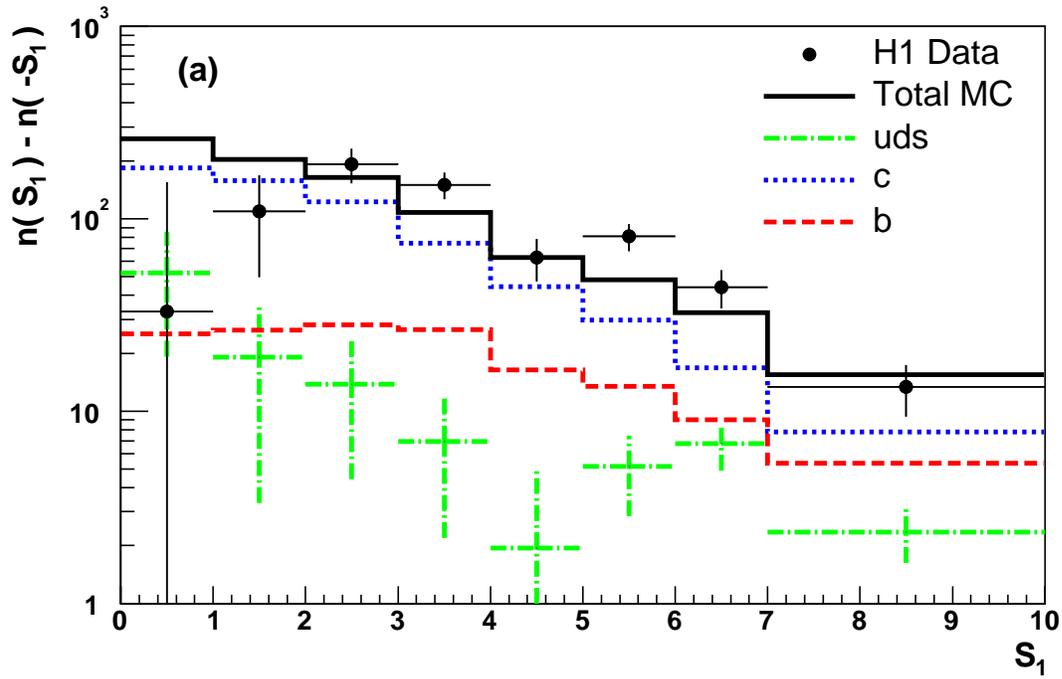}
    \includegraphics[width=1.0\textwidth]{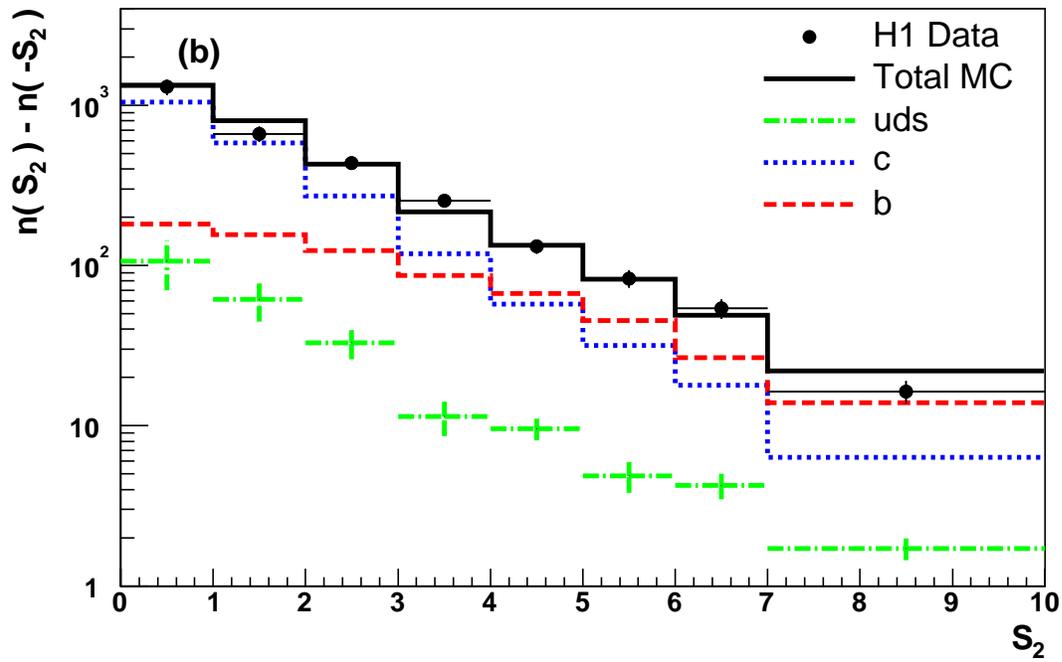} \caption{The subtracted
      distributions of (a) $S_1$ and (b) $S_2$.  Included in the figure is
      the result from the fit to the data of the Monte Carlo distributions
      of the various quark flavours.}
  \end{center}
  \label{fig:s1negsub} 
\end{figure}

The fractions of $c$, $b$ and light quarks of the data
are extracted in each $x$--$Q^2$ interval
using a  least squares simultaneous fit to 
the subtracted $S_1$ and $S_2$ distributions 
(as in figure~\ref{fig:s1negsub}) and
the total number of inclusive events before track selection.
The $c$, $b$ and $uds$ Monte Carlo simulation samples are used as templates.
Only the statistical errors of the data and Monte Carlo simulation are 
considered in the fit.  
The Monte
Carlo $c$, $b$ and $uds$ contributions in each $x$--$Q^2$ interval 
are allowed to be  scaled by factors $P_c$, $P_b$ and $P_l$, 
respectively. The fit to the $S_1$ and $S_2$ distributions mainly constrains
$P_c$ and $P_b$, whereas the overall normalisation constrains $P_l$.
The $c$ and $b$ quark fractions are distinguished in the fit by 
their different shapes in the $S_1$ and $S_2$ distributions. 

The results of the fit to the complete data sample are shown in
figure~\ref{fig:s1negsub}. The fit gives a reasonable description of
the $S_1$ distribution and a good description of $S_2$ distribution,
with a $\chi^2/ n.d.f$ of $27.5/14$.  Values of $P_c=0.81 \pm 0.08$, 
$P_b=1.62 \pm 0.24$
and $P_l=1.05 \pm 0.02$ are obtained.  Acceptable $\chi^2$ values are also
found for the fits to the samples in the separate $x$--$Q^2$
intervals.

Consistent results
are found when fitting
different significance distributions, for example fitting the $S_1$ or
$S_2$ distributions alone; fitting the highest absolute significance
track distribution for all events; fitting the distribution for
the track with the third highest absolute significance,
fitting the significance distributions
without subtraction of the negative bins from the positive, 
and also when varying the range of significance to be fitted. 
The analysis was also repeated excluding the CST tracks 
associated to the jet axis
from the primary vertex fit and compatible results were again found.

The results of the fit in each  $x$--$Q^2$ interval are converted to a 
measurement of the differential $c$ cross section using: 
\begin{equation}
\frac{{\rm d}^2\sigma^{c\bar{c}}}{{\rm d} x{\rm d} Q^2} = 
\frac{{\rm d}^2\sigma}{{\rm d} x{\rm d} Q^2} 
\frac{P_c N^{\rm MC gen}_c}{P_c N^{\rm MC gen}_c+P_b N^{\rm MC gen}_b+P_l N^{\rm MC gen}_l}  
\delta_{\rm BCC}^{c\bar{c}},
\end{equation}
where ${\rm d}^2 \sigma / {\rm d} x {\rm d} Q^2$ is the measured
inclusive differential cross section from H1\cite{H19900NCCC} and
$N^{\rm MC gen}_c$, $N^{\rm MC gen}_b$ and $N^{\rm MC gen}_l$ are the
generated number of $c$, $b$ and light quark events from the Monte
Carlo in each bin, respectively. A small bin centre correction
$\delta_{\rm BCC}^{c\bar{c}}$ is applied using the NLO QCD expectation (see
section~\ref{results}) to convert the bin averaged measurement into a
measurement at a single $x$--$Q^2$ point.  The cross section is
defined so as to include a correction for pure QED initial and final
state radiative effects, but not electroweak corrections (see
\cite{H19900NCCC} for a more complete discussion). Events that contain
$c$ hadrons via the decay of $b$ hadrons are not included in the
definition of the $c$ cross section.
The structure function $F_2^{c\bar{c}}$ is then evaluated
from the expression 
\begin{equation}
\frac{{\rm d}^2\sigma^{c\bar{c}} }{{\rm d} x{\rm d} Q^2} = \frac {2 \pi \alpha^2}{x Q^4 }  [(1+ (1-y)^2) F_2^{c\bar{c}}   - y^2  F_L^{c\bar{c}}],
\label{eq:sigcc}
\end{equation}
where the longitudinal structure function $F_L^{c\bar{c}}$ is
estimated from the NLO QCD expectation\cite{H19900NCCC}.
In the evaluation of $F_2^{c\bar{c}}$ the electroweak corrections are assumed to be
small and are neglected.
It is also convenient to express the cross section as a `reduced cross
section' defined as
\begin{equation}
\tilde{\sigma}^{c\bar{c}} (x, Q^2) = \frac{{\rm d}^2\sigma^{c\bar{c}} }{{\rm d} x{\rm d} Q^2}  \frac {x Q^4 } {2 \pi \alpha^2 (1+ (1-y)^2)}
= F_2^{c\bar{c}}   - \frac{y^2}{1+ (1-y)^2}  F_L^{c\bar{c}}.
\end{equation}
The differential $b$ cross section and $F_2^{b\bar{b}}$ are evaluated
in the same manner.

\subsection{Systematic Errors}
\label{systematics}
\label{sec:systematics}
The systematic uncertainties on the measured cross sections are
estimated by applying the following variations to the Monte Carlo
simulation:
\begin{itemize}
\item A  track efficiency uncertainty of $3\%$ due to the CTD and
of $2\%$ due to the CST.
\item An uncertainty in the $\delta$ resolution of the tracks (figure~\ref{fig:dca}(a))
is estimated by varying the resolution by an amount that encompasses
the differences between the data and simulation.  An additional
Gaussian smearing of $200$~$\mu{\rm m}$ to $5\%$ of randomly selected
tracks and $25$~$\mu{\rm m}$ to the rest is used.
\item A $4\%$ uncertainty on the hadronic energy scale.
\item An error on the jet axis is estimated by introducing an
  additional Gaussian smearing of 2$^\circ$ in azimuth.
\item The uncertainty on the asymmetry of the light quark $\delta$ 
  distribution is
  estimated by repeating the fits with the light quark $S_1$ and $S_2$
  distributions (figure~\ref{fig:s1negsub}) set to zero and doubling
  the contribution. 
  This error includes the modelling of light quark multiplicities.
\item The uncertainties on the various $D$ and $B$ meson lifetimes,
  decay branching fractions and mean charge multiplicities are
  estimated by varying the input values of the Monte Carlo simulation
  by the errors on the world average measurements, or by adjusting the
  simulation to the world average value depending on which variation
  is larger. For the branching fractions of $b$ quarks to hadrons and
  the lifetimes of the $D$ and $B$ mesons the central values and errors on the
  world averages are taken from\cite{Hagiwara:fs}.  For the branching
  fractions of $c$ quarks to hadrons the values and uncertainties are
  taken from\cite{Gladilin:1999pj}.  For the mean charged track
  multiplicities the values and uncertainties for $c$ and $b$ quarks
  are taken from MarkIII\cite{Coffman:1991ud} and
  LEP/SLD\cite{lepjetmulti} measurements, respectively.
\item An uncertainty on the fragmentation function of the heavy quarks
  is estimated using the Peterson fragmentation
  function\cite{peterson} with parameters $\epsilon_c = 0.058$ and
  $\epsilon_b = 0.0069$.
\item An uncertainty in the QCD model of heavy quark production
  is estimated by replacing the default RAPGAP model 
  (where heavy quarks are generated with only the PGF process)
  with RAPGAP used with a  $1:1$ ratio of QPM to PGF induced events.
\end{itemize}

Other sources of systematic error pertaining to the NC selection are
also considered\cite{H19900NCCC}: a $1.5\%$ uncertainty on the
luminosity measurement; an uncertainty on the scattered positron polar
angle of $1$--$3$~${\rm mrad}$ and energy of $0.7$--$3.0\%$ depending
on the polar angle; a $0.5\%$ uncertainty on the scattered positron
identification efficiency; a $0.5\%$ uncertainty on the positron
track-cluster link efficiency; a $0.3\%$ uncertainty on the trigger
efficiency and a $1\%$ uncertainty on the cross section evaluation due
to QED radiative corrections. An uncertainty due the bin centre
correction is estimated to be $5\%$. This corresponds to the maximum
correction for any $x$--$Q^2$ interval.

The uncertainties that contribute most to the total systematic error
on $\tilde{\sigma}^{c\bar{c}}$ are the track resolution and the light
quark $\delta$ asymmetry leading to errors on the cross section of $9\%$
each. Those that contribute most to the total systematic error on
$\tilde{\sigma}^{b\bar{b}}$ are the track resolution, the track
efficiency, and the QCD model leading to errors on the cross section
of $14\%$, $11\%$ and $8\%$, respectively.  The total systematic error
is obtained by adding all individual contributions in quadrature and
is around $15\%$ for $\tilde{\sigma}^{c\bar{c}}$ and $24\%$ for
$\tilde{\sigma}^{b\bar{b}}$.  The same systematic error uncertainty is
assigned to each of the four differential measurements. 

\subsection{Measurement Using Secondary Vertex Reconstruction}
\label{sec:secvertex}
The results are checked using an alternative method to separate the
quark flavours based on the explicit reconstruction of decay vertices
in the transverse plane.  In this approach, there is no definite
assignment of tracks to vertices, but each track is assigned a weight
with a range $0$ to $1$ for each vertex candidate, using the weight
function of \cite{ref:adaptive}.  The larger the distance of the track
to a vertex candidate, the smaller the weight.  A simultaneous fit to
a primary and a secondary vertex is made, with all tracks of the event
considered for the primary vertex, whereas only tracks associated to the
jet axis contribute to the secondary vertex.
The vertex configuration that minimises the global fit $\chi^2$ is
found iteratively using deterministic annealing\cite{ref:annealing}.

\begin{figure}
  \begin{center}
    \includegraphics[width=1.0\textwidth]{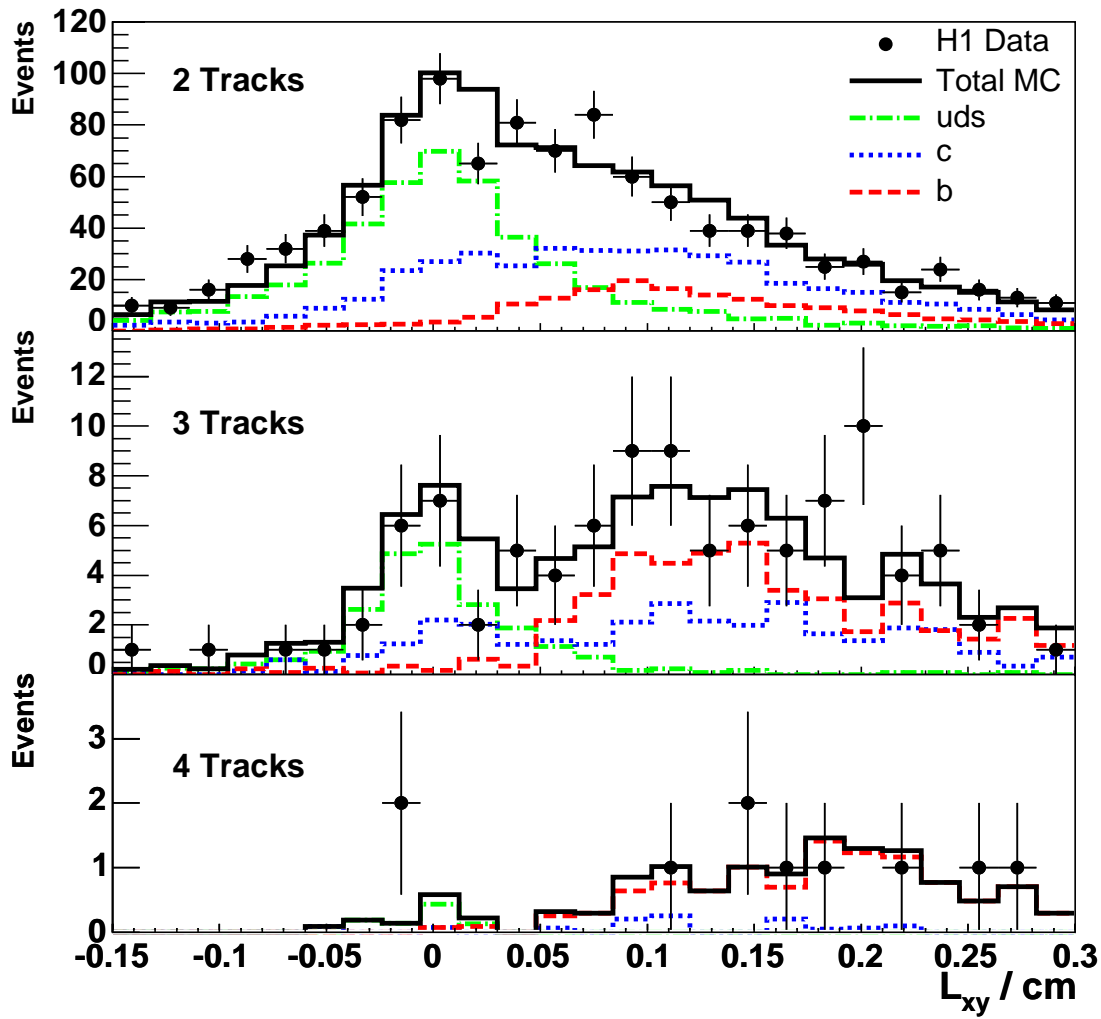}
    \caption{The transverse distance between
      the primary and secondary vertex ($L_{xy}$) for events
      with  two,  three and  four CST tracks associated with the
      secondary vertex. Superimposed on the data points are
      $c$, $b$ and light quark contributions that have been scaled
      by the results of the fit to the $S_1$ and
      $S_2$ data distributions.\label{fig:sl}
      }
  \end{center}
\end{figure}

In the secondary vertex reconstruction analysis the same event,
track and jet selections as applied in the impact parameter analysis are used.
The number of tracks contributing with a weight greater than $0.8$ to
the secondary vertex, after the last annealing step, is used as a
measure of the decay-multiplicity.  The transverse distance between
the primary and secondary vertex distributions $L_{xy}$ for different
decay-multiplicities is shown in figure \ref{fig:sl}.  The vertices
found in light quark events peak at $L_{xy} \simeq 0$, while the
vertices found in heavy quark events are significantly displaced in
the direction of the jet axis.  Both charm and beauty decays
contribute to the two track secondary vertex, whereas beauty dominates the 
three and four track secondary vertices.

When the $L_{xy}$ distributions for the different multiplicities are fitted
simultaneously using Monte Carlo templates for the $c$, $b$ and
$uds$ quark contributions to obtain the scale factors   
$P_c$, $P_b$ and $P_l$, the results are found to agree well with the 
impact parameter method. 
To illustrate this, the Monte Carlo contributions in 
figure~\ref{fig:sl} are scaled by the factors obtained
using the impact parameter method.

\section{Results}
\label{results}
\begin{figure}[htb]
  \begin{center}
    \includegraphics[width=\textwidth]{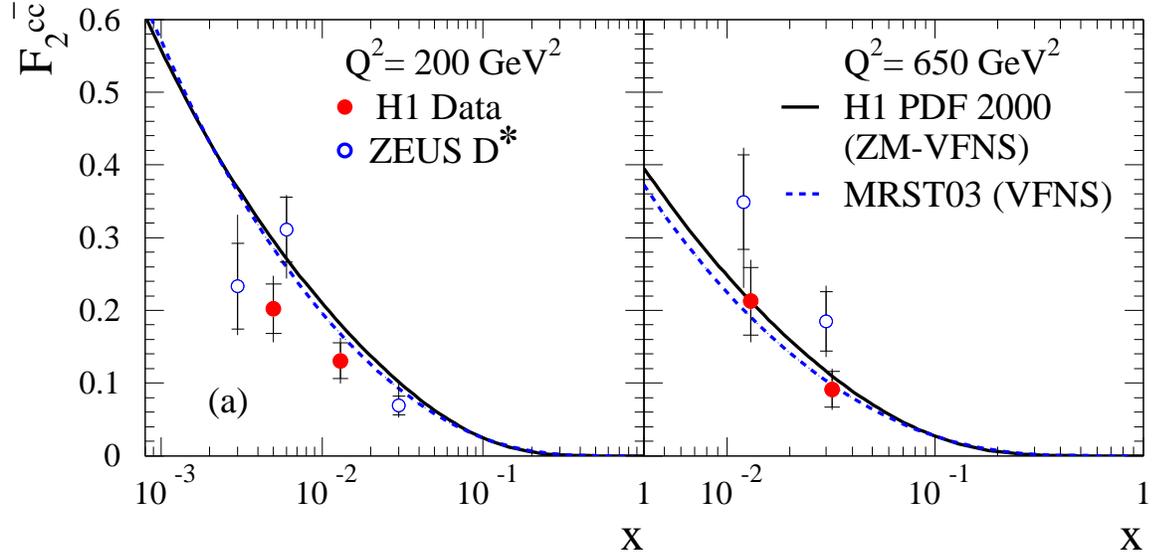}
    \includegraphics[width=\textwidth]{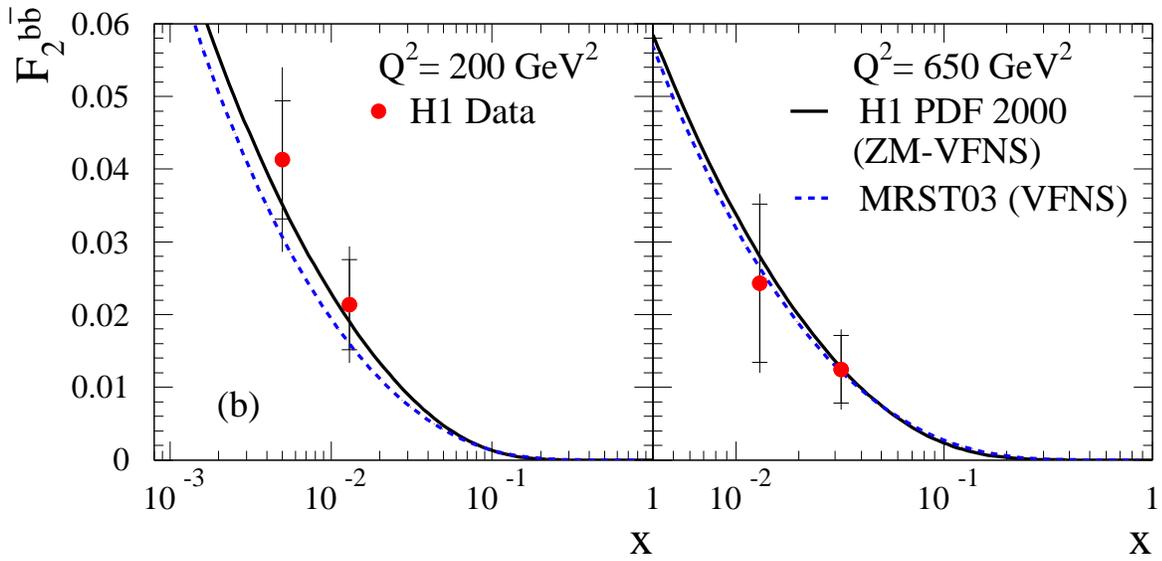}
    \caption{The measured $F_2^{c\bar{c}}$ (a) and $F_2^{b\bar{b}}$ (b)
      shown as a function of $x$ for two different $Q^2$ values.
      The inner error bars show the statistical error, the outer
      error bars represent the statistical and systematic errors 
      added in quadrature.
      The $F_2^{c\bar{c}}$ from ZEUS obtained from measurements
      of $D^*$ 
      mesons\cite{Chekanov:2003rb} and the predictions of 
      NLO QCD fits\cite{H19900NCCC, Martin:2003sk} are also shown.}
    \label{fig:f2c} 
  \end{center}
\end{figure}

The measurements of $F_2^{c\bar{c}}$ and $F_2^{b\bar{b}}$ are listed in
 table~\ref{tab:sig}
and shown in
figure~\ref{fig:f2c} as a function of $x$ for two
values of $Q^2$. 
The H1 data for $F_2^{c\bar{c}}$ are compared with the
results of the ZEUS collaboration\cite{Chekanov:2003rb} 
where the cross sections were
obtained from the measurement of $D^{*\pm}$ mesons. 
The results of the two measurements for $F_2^{c\bar{c}}$ are in good
agreement.

The data are also compared with two example predictions from 
NLO QCD (see section~\ref{sec:theory}).
These are the H1 PDF 2000 fit\cite{H19900NCCC} in which the $c$ and
$b$ quarks are treated in the ZM-VFNS scheme, 
and a fit from MRST03\cite{Martin:2003sk} which uses a VFNS scheme.  The
predictions of the two QCD approaches are similar
and compatible with the data.

\begin{figure}[htb]
  \begin{center} 
    \includegraphics[width=\textwidth]{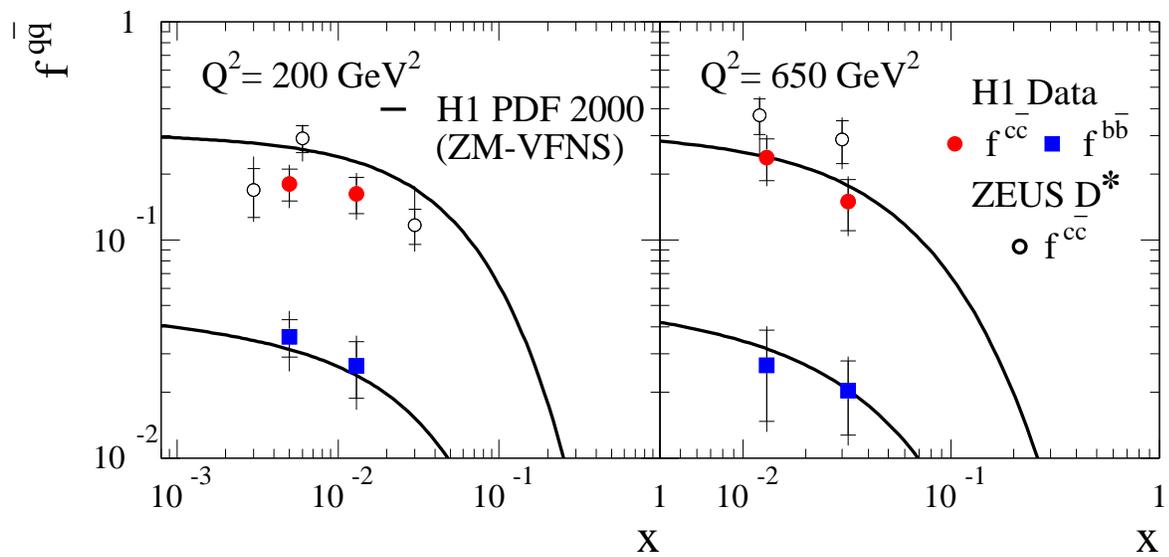} \caption{The
      contributions to the total cross section $f^{c\bar{c}}$ and
      $f^{b\bar{b}}$ shown as a function of $x$ for two different $Q^2$
      values. The inner error bars show the statistical error, the outer
      error bars represent the statistical and systematic errors added in
      quadrature.  The $f^{c\bar{c}}$ from ZEUS obtained from measurements
      of $D^*$ mesons\cite{Chekanov:2003rb} and the prediction of the H1
      NLO QCD fit\cite{H19900NCCC} are also shown.}  \label{fig:fraccb} 
  \end{center}
\end{figure}

The measurements are also presented in table~\ref{tab:frac}
and figure~\ref{fig:fraccb} 
in the form of the fractional contribution to the total $ep$ cross
section
\begin{equation}
f^{c\bar{c}} =  \frac{{\rm d}^2 \sigma^{c\bar{c}}} {{\rm d} x {\rm d} Q^2}
/
\frac{
{\rm d}^2 \sigma}{ {\rm d} x {\rm d} Q^2
}.
\end{equation}
The $b$ fraction $f^{b\bar{b}}$ is defined in the same manner.  NLO
QCD is found to give a good description of the data, as shown by
comparison with the ZM-VFNS prediction from the H1
PDF 2000 fit.

The $c$ and $b$ fractions and cross sections are also measured
integrated over the range \linebreak $Q^2>150~{\rm GeV}^2$  and $0.1<y<0.7$.
This is a more restricted range than for the differential measurements
in order that the acceptance for the scattered positron and products
of the $b$ and $c$ quarks is above $95$\%, integrated across the kinematic
range. The following values are found: 
\begin{equation*}
 f^{c\bar{c}} =  0.183 \pm 0.019 \pm 0.023, \,\, \,\,
 \sigma^{c\bar{c}} =  373 \pm 39 \pm 47 \ {\rm pb},
\end{equation*}
\begin{equation*}
 f^{b\bar{b}} =  0.0272 \pm 0.0043 \pm 0.0060, \,\, \,\,
 \sigma^{b\bar{b}} = 55.4 \pm 8.7 \pm 12.0 \ {\rm pb}.
\end{equation*}
The integrated cross sections may also be compared with the
predictions from NLO QCD.  The VFNS prediction from MRST03 gives
$\sigma^{c\bar{c}} = 426~{\rm pb}$ and $\sigma^{b\bar{b}}= 47~{\rm pb}$; the H1
PDF 2000 fit gives $\sigma^{c\bar{c}} = 455~{\rm pb}$ and
$\sigma^{b\bar{b}}= 52~{\rm pb}$.

It is also useful to compare with results from the FFNS scheme, which
was used for the QCD predictions in\cite{zeusBdis}.  Using the PDF
set CTEQ5F3\cite{Lai:1999wy} gives $ \sigma^{c\bar{c}} = 419 \ {\rm
  pb}$ and $\sigma^{b\bar{b}} = 37 \ {\rm pb}$. The values $m_c =
1.3~{\rm GeV}$ and $m_b = 4.75~{\rm GeV}$ were used and the
renormalisation and factorisation scales were set to $\mu = \sqrt{
  p^2_{T q\bar{q}} + m_b^2}$, where $p_{T q\bar{q}}$ is the mean
transverse momentum of the heavy quark pair.
Predictions for the cross sections may also be obtained from
fits\cite{ccfm} to the HERA inclusive $F_2$ data based on CCFM
evolution\cite{ccfm2}.  The CCFM predictions agree with those from the VFNS
prediction of MRST03 to within $7$\%.

All the QCD predictions are observed to be compatible with the data.
The errors on the data do not yet allow the different schemes to be
distinguished.  There is no evidence for a large excess of the $b$
cross section compared with QCD predictions as has been reported in
other measurements, which have been made in different kinematic ranges
or for different processes\cite{zeusBdis,Adloff:1999nr,hadronb,ggb}.

\section{Conclusion}

The inclusive charm and beauty cross sections in deep inelastic
scattering are measured at high $Q^2$ using a technique based on the
lifetime of the heavy quark hadrons. The measurements are made using
all events containing tracks with vertex detector information. In the
kinematic range of the measurements this eliminates the need for large
model dependent extrapolations to the full cross section.  Based on
impact parameter measurements in the transverse plane both integrated
and differential $c$ and $b$ cross sections are obtained.  The results
are verified using a method based on the explicit reconstruction of
decay vertices. The cross sections and derived structure functions
$F_2^{c\bar{c}}$ and $F_2^{b\bar{b}}$ are found to be well described
by predictions of perturbative QCD. This is the first measurement of
$F_2^{b\bar{b}}$.

\section*{Acknowledgements}

We are grateful to the HERA machine group whose outstanding efforts
have made this experiment possible.  We thank the engineers and
technicians for their work in constructing and maintaining the H1
detector, our funding agencies for financial support, the DESY
technical staff for continual assistance and the DESY directorate for
support and for the hospitality which they extend to the non-DESY
members of the collaboration.


\newpage

\begin{table}
 \begin{tabular}{|c|c|c||c|c|c|c||c|c|c|c|} \hline 
 $x$ & $y$ & $Q^2$ & $\tilde{\sigma}^{c\bar{c}}$ & $\delta^{c\bar{c}}_{\rm stat}$ & $\delta^{c\bar{c}}_{\rm sys}$ & $F_2^{c\bar{c}}$ &  $\tilde{\sigma}^{b\bar{b}}$ & $\delta^{b\bar{b}}_{\rm stat}$ & $\delta^{b\bar{b}}_{\rm sys}$ & $F_2^{b\bar{b}}$ \bigstrut[t] \\ 
 & & (GeV$^2$) & & (\%) &  (\%) & & & (\%) &  (\%) & \\ \hline
  0.0050  &   0.394   &  200 & 0.197  &    17   &  15  &   0.202   &  0.0393  &    20   &  24  &    0.0413  \\  
  0.0130  &   0.151   &  200 & 0.130  &    19   &  15  &   0.131   &  0.0212  &    29   &  24  &    0.0214  \\  
  0.0130  &   0.492   &  650 & 0.206  &    22   &  15  &   0.213   &  0.0230  &    45   &  24  &    0.0243  \\  
  0.0320  &   0.200   &  650 & 0.091  &    27   &  15  &   0.092   &  0.0124  &    37   &  24  &    0.0125  \\  
 \hline 
 \end{tabular}

    \caption{ 
The measured reduced NC  charm ($\tilde{\sigma}^{c\bar{c}}$) and beauty
($\tilde{\sigma}^{b\bar{b}}$) cross sections, shown with statistical
 ($\delta^{c\bar{c}}_{\rm stat}$,
 $\delta^{b\bar{b}}_{\rm stat}$) and systematic 
($\delta^{c\bar{c}}_{\rm sys}$,  
$\delta^{b\bar{b}}_{\rm sys}$) errors.
The table also shows the values for $F_2^{c\bar{c}}$ and $F_2^{b\bar{b}}$
obtained from the measured cross sections using the NLO QCD fit to correct
for $F_L^{c\bar{c}}$ and  $F_L^{b\bar{b}}$.
}
\label{tab:sig} 
\end{table}

\begin{table}
 \begin{tabular}{|c|c|c||c|c|c||c|c|c|} \hline 
 $x$ & $y$ & $Q^2$ & $f^{c\bar{c}}$ & $\delta^{c\bar{c}}_{\rm stat}$ & $\delta^{c\bar{c}}_{ \rm sys}$ &  $f^{b\bar{b}}$ & $\delta^{b\bar{b}}_{\rm stat}$ & $\delta^{b\bar{b}}_{\rm sys}$   \bigstrut[t] \\ 
 & & (GeV$^2$) & & (\%) &  (\%) &  & (\%) &  (\%)  \\ \hline
  0.0050  &   0.394   &  200 & 0.181  &    17   &  15      &  0.0361  &    20   &  24      \\  
  0.0130  &   0.151   &  200 & 0.163  &    19   &  15      &  0.0265  &    29   &  24      \\  
  0.0130  &   0.492   &  650 & 0.239  &    22   &  15      &  0.0266  &    45   &  24      \\  
  0.0320  &   0.200   &  650 & 0.150  &    27   &  15      &  0.0203  &    37   &  24      \\  
 \hline 
 \end{tabular}  

    \caption{ The measured charm ($f^{c\bar{c}}$) and beauty
($f^{b\bar{b}}$) fractional contributions to
the total cross section, shown with 
statistical ($\delta^{c\bar{c}}_{\rm stat}$, 
$\delta^{b\bar{b}}_{\rm stat}$) and systematic ($\delta^{c\bar{c}}_{\rm sys}$,  
$\delta^{b\bar{b}}_{\rm sys}$) errors.}
\label{tab:frac} 
\end{table}

\newpage

\end{document}